# Few-molecule reservoir computing experimentally demonstrated with surface enhanced Raman scattering and ion-gating stimulation

Daiki Nishioka[1,2], Yoshitaka Shingaya[1], Takashi Tsuchiya[1]*, Tohru Higuchi[2], Kazuya Terabe[1]

[1]Research Center for Materials Nanoarchitectonics (MANA), National Institute for Materials Science (NIMS), 1-1 Namiki, Tsukuba, Ibaraki 305-0044, Japan.
[2]Department of Applied Physics, Faculty of Science, Tokyo University of Science, 6-3-1 Niijuku, Katsushika, Tokyo 125-8585, Japan.

Email: TSUCHIYA.Takashi@nims.go.jp

**Abstract**

Reservoir computing (RC) is a promising solution for achieving low power consumption neuromorphic computing, although the large volume of the physical reservoirs reported to date has been a serious drawback in their practical application. Here, we report the development of a few-molecule RC that employs the molecular vibration dynamics in the para-mercaptobenzoic acid (pMBA) detected by surface enhanced Raman scattering (SERS) with tungsten oxide nanorod/silver nanoparticles ($WO_x$@Ag-NPs). The Raman signals of the pMBA molecules, adsorbed at the SERS active site of $WO_x$@Ag-NPs, were reversibly perturbated by the application of voltage-induced local pH changes in the vicinity of the molecules, and then used to perform RC of pattern recognition and prediction tasks. In spite of the small number of molecules employed, our system achieved good performance, including 95.1% to 97.7% accuracy in various nonlinear waveform transformations and 94.3% accuracy in solving a second-order nonlinear dynamic equation task. Our work provides a new concept of molecular computing with practical computation capabilities.

# Introduction

While AI provides convenience in variety of human activities through high performance deep learning, the huge energy consumption and heavy communication traffic that accompanies the current cloud-based AI systems have become serious problems [1]. Recently, in an effort to overcome the drawbacks of current AI systems, physical reservoir computing (PRC) has begun to attract attention [2,3]. Conventional deep learning requires massive computation resources to handle the enormous networks they require [i.e., vast numbers of weights, stored and updated in the learning (training) process], while PRC is theoretically predicted to have comparable performance due to its harnessing of the inherent nonlinear responses of physical dynamical systems, termed 'physical reservoirs' in the framework of reservoir computing (RC), for mapping time-series data to high dimensional feature space in order to achieve efficient information processing [4,5]. To the present, in the search for high performance PRC, lots of materials and devices (e.g., memristor, spintronics devices, soft bodies, nanowire networks, optical circuits, ionic devices) have been intensively explored and applied to a variety of pattern recognition and prediction tasks, including nonlinear transformation tasks, handwritten digit recognition, and face classification [2,3,5-36]. While good computational performance has been demonstrated in some cases, making PRC particularly attractive for next generation leading-edge AI systems, one serious problem has arisen; notably the larger volume of physical reservoirs compared to the volumes found in semiconductor elements/arrays [2,3,12-32]. One typical example is nonlinear optical circuits, which require bulky optical fiber in lengths as long as several km to perform PRC [18-20]. In this regard, a small number of molecules, with an extremely small volume, is a promising candidate for the realization of high performance and highly integrated PRC, which would lead to its implementation in electronic devices. The concept of molecular reservoir computing was previously proposed, and there have been a few reports of work on theoretical simulations and experimental demonstrations with preliminary tasks [24-36]. However, the number of molecules in the reported works was huge in each case (e.g., a network of single-walled carbon nanotubes (SWNTs) complexed with polyoxometalate (POM) [24,25], organic dendritic networks composed of the semiconductor PEDOT:PF6 [26,27], a sulfonated polyaniline network [28], a p-NDI transistor [29], an electrochemical cell using ionic liquid [30,31] and a POM molecule in the solution [32], a Polymerase-Exonuclease-Nickase Dynamic Network Assembly toolbox [33], a gene regulation Network with mRNA [34], a deoxyribozyme oscillator with DNA molecules [35]). Therefore, to date, the feasibility of single to few molecular reservoir computing with a practical tasks is unclear [35]. In particular, it is not straightforward to track the nonlinear dynamics of few molecules sufficiently precisely to enable PRC.

Here, we demonstrate few-molecule reservoir computing (FM-RC) using the molecular vibration dynamics in para-mercaptobenzoic acid (pMBA). The molecular dynamics of a few pMBA molecules was detected by surface enhanced Raman scattering (SERS) with tungsten oxide nanorod/silver nanoparticles ($WO_x$@Ag-NPs) [37]. In order to perform RC of standard benchmark tasks, the Raman signal of the pMBA molecules, adsorbed at the SERS active site of $WO_x$@Ag-NPs,

was reversibly perturbated by applied voltage-induced local pH changes in the vicinity of the molecules [38-42]. In spite of only a few molecules being present, our system achieved good performance accuracy (95.1% to 97.7%) in various nonlinear waveform transformation and 94.3% accuracy in a second order nonlinear transformation task. Our results open a solid new direction for molecular computing by combing excellent compatibility with high integration, low power consumption, and a practical computational capacity for pattern recognition and prediction tasks [35,43-48].

**Results and Discussion**

The molecular vibrations of few molecules were detected using the very large surface enhanced Raman scattering (SERS) effect of conductive tungsten oxide ($WO_x$) [37]. Figure 1a shows a schematic diagram of the experimental arrangement for the subject Raman scattering measurements, in an electrochemical system. A single $WO_x$ nanorod was fixed, with epoxy resin, to the apex of a sharpened tungsten tip. The tip was inserted into the meniscus of the electrolyte solution formed between a 100× immersion objective lens and a petri dish, following which a laser beam was irradiated to the center of the $WO_x$ nanorod. Voltage was applied to the $WO_x$ nanorod in the electrochemical arrangement shown in Fig. 1a. The tungsten tipped $WO_x$ nanorod was used as the working electrode, and a platinum wire was used as the counter and pseudo-reference electrode. Figure 1b is an optical microscope image of the $WO_x$ nanorods observed in an aqueous electrolyte solution. The $WO_x$ nanorods used here have a {001} type crystallographic shear (CS) structure in the crystal, and exhibit metallic electrical conduction properties [49-51]. One of the authors previously reported that $WO_x$ nanorods with this type of crystalline structure exhibit a very large Raman enhancement effect, and can be used for single-molecule Raman detection [37]. It is thought that a large enhancement effect can be obtained by introducing defects into the crystallographic shear structure and forming a nanogap on the conducting plane. In this study, in addition to introducing such defects on the conducting plane of the $WO_x$ nanorods, a small number of silver nanoparticles were attached to the $WO_x$ nanorods to form a nanogap between the $WO_x$ nanorods and the silver nanoparticles, so as to obtain an additional enhancement effect. This method is suitable for observing Raman scattering of adsorbed molecules from electrolyte solutions because it more reliably forms a SERS active site on the $WO_x$ nanorod surface. In Figure 1b, the central area of the $WO_x$ nanorods appears brighter. Silver nanoparticles are electrodeposited in this region and form a SERS active site. Since electrodeposition on $WO_x$ nanorods occurs at the light-irradiated area, it is possible to control the position at which the silver particles are attached. Figure 1c is a transmission electron microscope (TEM) image of tungsten oxide nanorods in the area where the silver particles were electrodeposited. It can be seen that the silver nanoparticles, 5—30 nm in diameter, have become attached to the $WO_x$ nanorod.

Figure 1d is the Raman scattering spectra of a silver nanoparticle-modified $WO_x$ nanorod, obtained in a 0.15 M NaCl solution with 10 µM p-mercaptobenzoic acid (pMBA). The structure of the pMBA molecule changes with the pH in the carboxyl groups, which difference is clearly shown in the Raman spectra and which makes it possible for it to be used for local pH sensing [38-42]. In the present study, changes in electrode potential provide spectral changes comparable to those obtained by changing pH. At +160 mV vs SHE, a ν(C=O) peak appears at 1706 cm$^{-1}$, indicating that the carboxyl groups of the adsorbed molecules are protonated. In contrast, the 1706 cm$^{-1}$ peak disappears completely at -340 mV vs SHE and a ν(COO$^-$) peak appears at 1397 cm$^{-1}$, which indicates that the carboxyl groups of pMBA are deprotonated. Figure 1e is a schematic diagram of vibrational modes corresponding to the peaks that appeared at 1077, 1586, 1706 cm$^{-1}$ of protonated pMBA and at 1077, 1397, 1582 cm$^{-1}$ of deprotonated pMBA. [52-54] The peaks at 1077 cm$^{-1}$ were assigned to a ring breathing mode and a ν(C-S) mode. The peaks at 1586 and 1582 cm$^{-1}$ were assigned to a ring stretching mode. FM-RC utilizes such potential-dependent molecular vibrational behavior of only a few molecules as a computational resource for information processing. Unlike conventional methods that use the macroscopic network structure and electrical properties formed by a vast number of organic molecules as a computational resource, our approach takes advantage of the information processing capability of the microscopic nonlinear behavior of a few molecules. Figure 1f shows the time-resolved SERS spectra of a silver nanoparticle modified WO$_x$ nanorod, obtained by reducing the concentration of pMBA to 1 nM. Characteristic blinking phenomena are exhibited when observing extremely small amounts of molecules at the single-molecule level. Therefore, it can be concluded that the present system, in which silver particles are electrodeposited on WO$_x$ nanorods, provides a very large enhancement effect that enables single molecule detection.

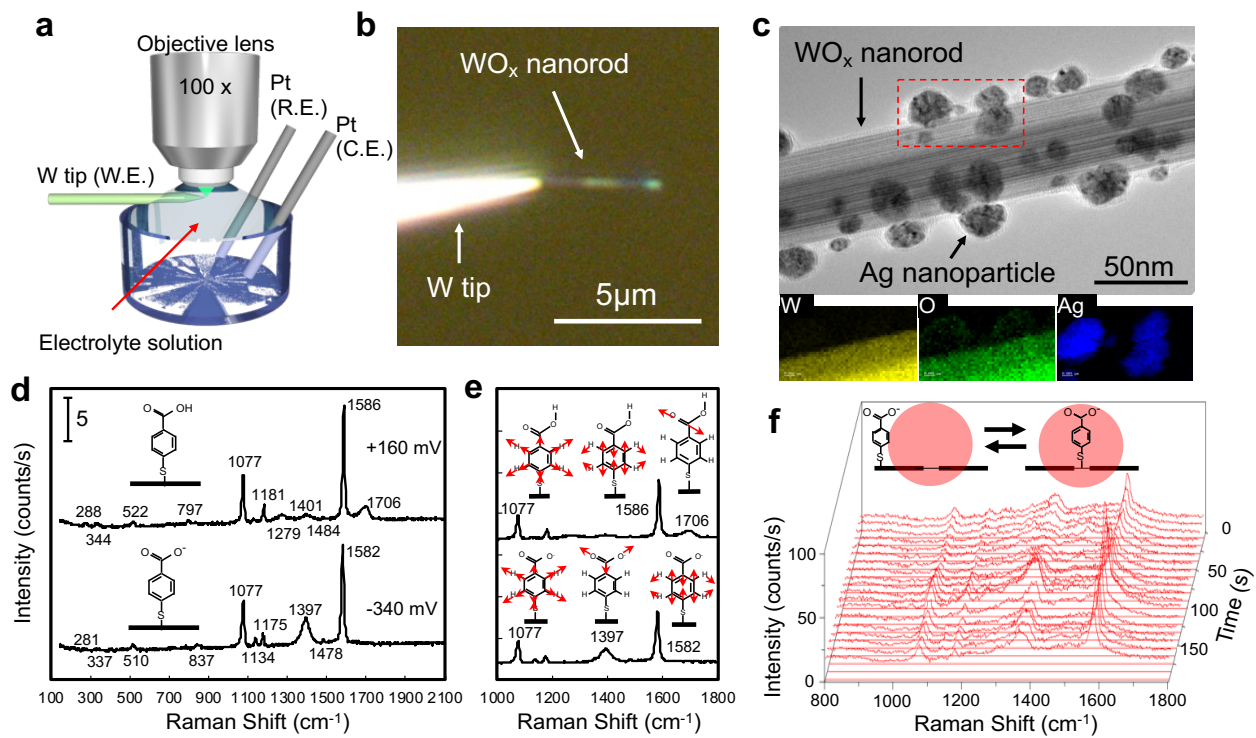

**Fig. 1 a** Schematic diagram of laser irradiation and electrode arrangement for SERS measurements. **b** Optical microscope image of a WO$_x$ nanorod attached at the apex of a tungsten tip. **c** TEM image of a WO$_x$ nanorod modified with silver nanoparticles. The three images at the bottom of the figure represent EDS maps of the area enclosed by the red dotted square in the TEM image. The three images represent EDS maps of tungsten, oxygen and silver, respectively. **d** SERS spectra obtained from Ag nanoparticle modified WO$_x$ nanorod in a 0.15 M NaCl aqueous solution. The pMBA concentration was 10 µM, the excitation laser wavelength was 488 nm, and the laser power was 20µW. **e** Schematic diagram of vibrational modes corresponding to the peaks appearing at 1077, 1586, 1706 cm$^{-1}$ of protonated pMBA and at 1077, 1397, 1582 cm$^{-1}$ of deprotonated pMBA. **f** Time-resolved SERS spectra obtained from WO$_x$ nanorod modified with silver nanoparticles. The pMBA concentration was 1 nM.

**Nonlinear waveform transformation task**

A nonlinear waveform transformation task [8,16,23] was performed to demonstrate the computational capability of the FM-RC. Figure 2a is a schematic of the task calculated by the FM-RC. The goal of this task is to generate sin wave (Sin), square wave (Square), π/2 phase-shifted triangular wave (Shift), and frequency-doubled triangular wave (2f) by the linear combination (Eq. 1) of readout weights $W$ and the reservoir states $X$ (in this case Raman spectrum) to the triangular wave input.

$$Y(k) = WX(k) \qquad (1)$$

, where $k$ and $Y(k)$ are the discrete time step and reservoir output, respectively. This task requires nonlinearity and high dimensionality in the reservoir, by which the basic computational capability of the reservoir can be evaluated. The input triangular wave for the nonlinear waveform transformation task was converted into a voltage signal of 10 s per step and 20 steps per period, and input to the WO$_x$ nanorod working electrode on which the pMBA molecule was adsorbed. This was done in the range of -340mV to +160 mV vs SHE. The Raman spectral response for time steps to the periodic triangular wave input is shown in Fig. 2b. It can be seen that the bright SERS bands including ν(COO$^-$) peak at 1384 cm$^{-1}$ and ν(C=O) peak at 1693 cm$^{-1}$ [38,39], and exhibit periodic blinking by responding to the triangular wave input. This is due to the applied voltage-induced reversible protonation and deprotonation of the carboxyl groups of pMBA, as shown in Fig. 1d. Two hundred reservoir states were obtained from these Raman spectra, uniformly in the range of 945 cm$^{-1}$ to 1728 cm$^{-1}$. This means that not only the reversible and independent nonlinear dynamics of the prominent peaks corresponding to the ring-breathing mode and the ν(C-S) mode (1077 cm$^{-1}$), the ring-stretching mode (1582~1586 cm$^{-1}$), but also the reversible and independent nonlinear dynamics of the minor structures shown in Fig. 1e corresponding to ν(COO$^-$) mode (1397 cm$^{-1}$) and ν(C=O) (1706 cm$^{-1}$) contribute to perform the task. The readout weights were trained by ridge regression so that the reservoir output $Y$(k), obtained by the linear combination of reservoir states and weights, agreed with the target waveform $T$(k). Details of the method and learning algorithm are given in the Methods section.

Figure 2c shows the target waveform and the output waveform from the molecule reservoir for each transformation task. For all tasks, the FM-RC generates waveforms that are in good agreement with the target waveforms, which means that the voltage response of only a few pMBA molecules has the capability of performing nonlinear transformations that can be used for information processing in reservoir computing. The subject FM-RC achieved 97.6% accuracy for 'Sin', 95.1% for 'Square', 97.7% for 'Shift', and 97.1% for '2f' in each conversion task. Whereas the performance for the 'Sin' task was comparable to other physical reservoirs, the subject FM-RC outperformed a YSZ/Diamond-based electric double layer-ion-gating reservoir (EDL-IGR) and nanowire network (NWN) reservoirs for the other tasks, as shown in Fig. 2d [8,16,23]. In particular, the FM-RC achieved excellent accuracy of as high as 97.1% for the 2f task, which task requires the strongest and most complex variety of nonlinearities among the four transformation tasks, which NWN reservoirs could thus only solve under extremely limited conditions that were tuned to an 'edge of chaos' state [16] that maximizes the information processing capabilities as a dynamical system [55,56]. In addition, for EDL-IGR [23], which performed the 2f task with the same high accuracy as FM-RC, the synaptic response based on ion-electron coupling dynamics was reported to be in an 'edge of chaos' state [21]. The fact that FM-RC outperformed such complexity exhibiting PRC, indicates the good nonlinearity and complexity of the subject FM-RC. In the nonlinear waveform transformation task, the extent of the nonlinearity in the physical reservoir dynamics is mainly reflected in the transformation accuracy. The present result evidences that reversible protonation and deprotonation of pMBA, caused by the local pH change, owns strong nonlinearity that is comparable to that exhibited by the electrical response of an atomic switch nanowire network, which is itself accompanied by collective resistive switching at numerous cross points in the network [8,16]. Furthermore, the result supports the contention that the SERS technique, utilizing a silver nanoparticle modified $WO_x$ nanorod, is capable of detecting the nonlinear response of few-molecule pMBA with a sufficient signal-to-noise ratio due to its excellent sensitivity. On the other hand, in addition to nonlinearity, reservoir computing requires short term memory and high dimensionality, and these have a relatively small impact on the accuracy of the task. To investigate the versatility of the molecular reservoir for physical reservoir computing tasks needing all the three requirements (i.e., nonlinearity, short term memory, high dimensionality), a more difficult task was performed. This is discussed in the following section.

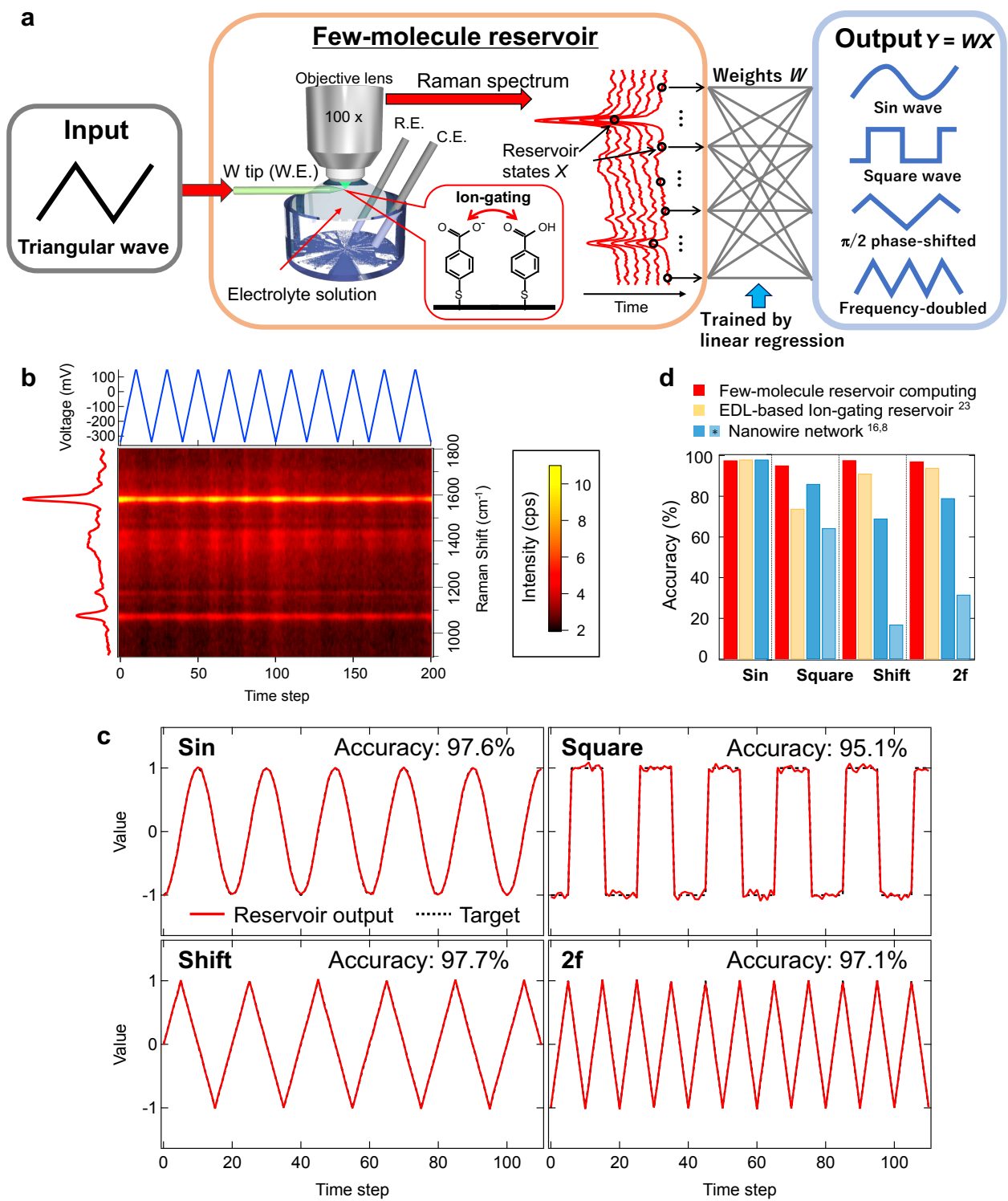

**Fig. 2 Nonlinear waveform transformation task. a** Schematic of the nonlinear waveform transformation task calculated by FM-RC. **b** Response of Raman spectra to triangular wave voltage application. **c** Results of the waveform transformation task using a sine wave, a square wave, a phase-shifted triangular wave, and a frequency-doubled triangular wave as the target waveform. The red line indicates the reservoir output and the black dotted line indicates the target waveform. **d** Comparison of the accuracy of the waveform transformation task with other physical reservoirs [8,16,23]. Results of NWN, with sin wave used as input indicated by * [8].

**Solving a second-order nonlinear dynamic equation task**

We evaluated the performance of the subject FM-RC with the more difficult task of solving a second-order nonlinear dynamic equation [5,11,21,22], since the previous waveform transformation task is relatively easy. In this task, in the training phase, the readout weights are trained so that the reservoir output matches the second-order nonlinear equation shown in Eq. 2. Then, it tests to make sure that the reservoir outputs (Eq. 1) are the same as the target (Eq. 2) for the untrained data (test phase),

$$T(k) = 0.4T(k-1) + 0.4T(k-1)T(k-2) + 0.6u^3(k) + 0.1 \qquad (2)$$

where $u(k)$ is a random input ranging from 0 to 0.5. The reservoir is required to have at least two steps of short-term memory and nonlinearity in order to express Eq. 2. A detailed explanation of the learning algorithm can be found in the Methods section. Figure 3b shows the reservoir output and the target waveform during the training phase. Both waveforms are in good agreement, and the molecule reservoir learned the second-order nonlinear equations with 94.5% accuracy. In the test phase, a data set different from the one used in training was input to the FM-RC, and its prediction output was compared to a target generated by Eq. 2. As shown in Fig. 3c, the target waveform and the predicted waveform of the FM-RC were in good agreement, even using a dataset that was different from the one used in training, and the FM-RC was able to predict the target waveform with an accuracy of 94.3%. Figure 3d shows the performance of the subject FM-RC compared with other physical reservoirs reported to date. Although the performance exhibited was not as good as that of spin torque oscillators (STO) [11] (theoretical calculation) and ion-gating reservoirs (IGRs) [21,22] which use semiconductor channels, the subject FM-RC, despite its extremely small reservoir volume, achieved similar computational performance as a memristor RC [5]. This indicates that the nonlinear dynamics reflecting structural changes caused by local pH changes in just a few pMBA molecules, measured via SERS measurements, has sufficient expressive power and short-term memory, as a dynamical system, to solve Eq. 2. As discussed above, RC computational performance depends on the nonlinearity and high dimensionality of the physical reservoir as a dynamical system in addition to these features. These characteristics were analyzed to investigate the origin of the performance, as described in the following section.

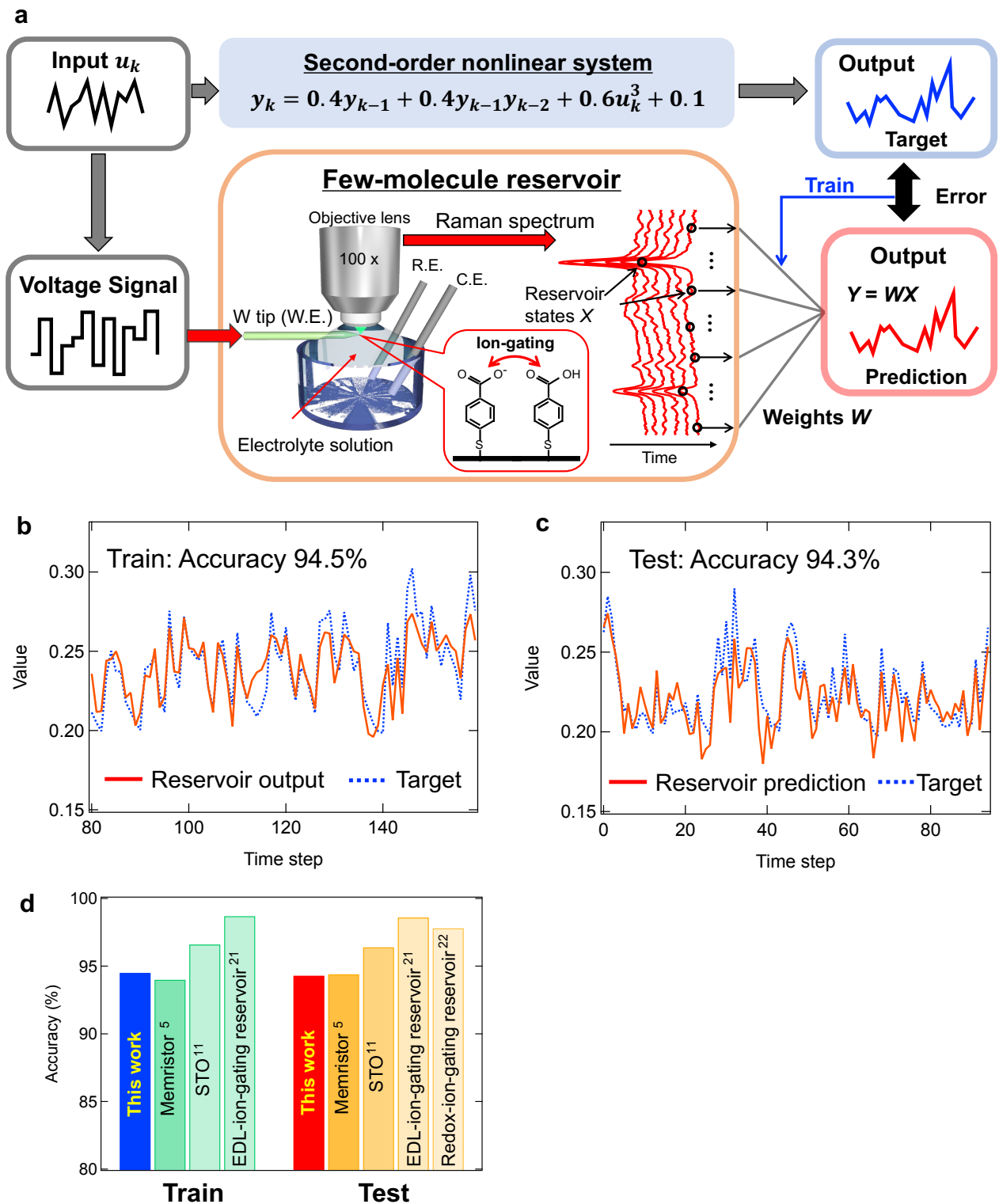

**Fig. 3 Solving a second-order nonlinear dynamic equation task. a** Schematic of task calculated by FM-RC. Target and prediction waveforms of second-order nonlinear dynamic equation at the (**b**) train phase and (**c**) test phase. **d** Prediction accuracy compared to other physical reservoirs (memristor, spin torque oscillator: STO, electric double layer type ion-gating reservoir: EDL-IGR, redox type ion-gating reservoir) [5,11,21,22].

**Nonlinearity and high dimensionality analysis of the few-molecule reservoir**

We analyzed the origin of the computational capability of the FM-RC in terms of nonlinearity, which is a required property for reservoirs. In reservoir computing, the nonlinearity of the reservoir is directly related to the nonlinear information processing capability, and strong nonlinearity with respect to the input contributes significantly to the expressive power of the reservoir. Figure 4a shows the input voltage dependence of the reservoir states obtained from 1571 cm$^{-1}$ and 1587 cm$^{-1}$. The input-output characteristics of the reservoir state at 1587 cm$^{-1}$ ($X_{1587cm^{-1}}$), corresponding to the peak top, are relatively linear, whereas the reservoir state at 1571 cm$^{-1}$ ($X_{1571cm^{-1}}$) changes nonlinearly. This is attributed to the ring-stretching peak not only increasing in intensity as the input voltage decreases from +160mV to -340mV vs SHE, but also shifting to the lower wavenumber side, as shown in the inset of Fig.4a. For quantitative evaluation of the nonlinearity of these reservoir states, the reservoir outputs for normalized input voltages $V_n$(=0 to -1) are shown in Fig. 4b, along with the input-output characteristics of a redox-IGR for comparison[22]. The nonlinearity of the reservoir state $X(V_n)$ is quantified by the correlation coefficient $r_{nonlinearity}$ (Eq. 3) with the linear line ($Y_{linear}=-V_n$): where the correlation coefficient between $Y_{linear}=-V_n$ and reservoir state $X(V_n)$ is 1, the input-output characteristics of that reservoir state are perfectly linear. On the other hand, as $r_{nonlinearity}$ gets closer to 0, the nonlinearity of the reservoir state becomes stronger.

$$r_{\text{nonlinearity}} = \frac{\sum (X(V_n) - \bar{X})(Y_{\text{linear}} - \bar{Y}_{\text{linear}})}{\sqrt{\sum (X(V_n) - \bar{X})^2}\sqrt{\sum (Y_{\text{linear}} - \bar{Y}_{\text{linear}})^2}} \quad (3)$$

, where $\bar{X}$ is the average of $X$. The redox-IGR, which utilizes the redox reaction by ion gating to WO$_3$ as the nonlinear dynamics, has $r_{nonlinearity}$ =0.964, which is close to 1 and the nonlinearity is weak. On the other hand, FM-RC, which utilizes the structural changes associated with local pH changes in the pMBA molecule as nonlinear dynamics, achieved a wide range of nonlinearities, from a strong nonlinearity ($r_{nonlinearity}$ =0.885) in the reservoir state obtained at 1571 cm$^{-1}$, to a weak nonlinearity ($r_{nonlinearity}$ =0.978) in the reservoir state obtained at 1587 cm$^{-1}$ nonlinearity. Such coexistence of reservoir states with different nonlinearities has been reported to improve performance in full-simulation reservoirs such as the echo state network (ESN) [57] and in physical reservoirs such as IGRs [21-23]. The excellent results in the nonlinear waveform transformation task are attributed to the good nonlinear characteristics of the FM-RC.

       High dimensionality is, in addition to nonlinearity, another important property required of a reservoir. High dimensionality refers to the diversity of reservoir states, which is achieved by having many reservoir states that exhibit different behaviors from each other. In simulation reservoirs such as ESN, high dimensionality can be easily achieved simply by increasing the network size (number of nodes), and although the computational cost increases, memory capacity and computational performance also increase as the network size increases [4]. On the other hand, in a physical reservoir, a sufficiently complex and diverse set of nodes based on the inherent nonlinear dynamics of the physical system is considered to inherently exist, but the number of effectively available nodes is not large due to the limited methods of accessing them (e.g., measurement probes). Therefore, to increase

the number of nodes, the method usually adopted is that of regarding the time evolution of reservoir states obtained from the physical system as spatially different nodes (virtual nodes) [9-12,15,18-23,58]. However, as the number of virtual nodes increases, the number of similar nodes increases, thus limiting the higher dimensionality. Here, for the Raman spectra of detected structural changes of pMBA molecules in response to local pH changes, the intensities of different wavenumbers were high-dimensionalized (without virtual nodes) as spatially distinct reservoir states. The correlation between reservoir state $X_i$ and reservoir state $X_j$ was quantified by the following correlation coefficient $r_{ij}$:

$$r_{ij} = \frac{\sum_{k=1}^{L}(X_i(k) - \bar{X}_i)(X_j(k) - \bar{X}_j)}{\sqrt{\sum_{k=1}^{L}(X_i(k) - \bar{X}_i)^2}\sqrt{\sum_{k=1}^{L}(X_j(k) - \bar{X}_j)^2}} \quad (4)$$

The lower panel of Fig. 4c shows the correlation coefficient $|r_{ij}|$ of the reservoir states of the subject FM-RC for random wave inputs. The darker areas in the figure indicate regions where the correlation coefficient is low and contributes to the high dimensionality of the reservoir, while the lighter areas in the figure indicate that the correlation coefficient is close to 1 and the reservoir states are similar to each other. To show how the reservoir states differ at these correlation coefficient values, the time evolution of the reservoir states is shown in Figure 4d as an example for a combination with relatively low correlation coefficients ($X_{1587\text{cm}^{-1}}$ and $X_{1287\text{cm}^{-1}}$) and a combination with correlation coefficients close to 1 ($X_{1587\text{cm}^{-1}}$ and $X_{1579\text{cm}^{-1}}$). $X_{1587\text{cm}^{-1}}$ and $X_{1287\text{cm}^{-1}}$ shown in the upper panel of Fig. 4d exhibit different behaviors from each other, whereas $X_{1587\text{cm}^{-1}}$ and $X_{1579\text{cm}^{-1}}$ shown in the lower panel exhibit almost the same behavior. A scatter plot of these reservoir states is also shown in Fig. 4e. The $X_{1587\text{cm}^{-1}}$ vs. $X_{1287\text{cm}^{-1}}$ plot shown in the upper panel shows that the correlation coefficient between them corresponds to $|r_{ij}|$=0.483, indicating that the correlation between them is relatively low, thus contributing to the high dimensionality of the reservoir. On the other hand, in the $X_{1587\text{cm}^{-1}}$ vs $X_{1579\text{cm}^{-1}}$ plot shown in the lower panel of Fig. 4e, the correlation coefficient between them corresponds to $|r_{ij}|$=0.971, which indicates that these reservoir states are almost completely correlated. In the $|r_{ij}|$ plot shown in the lower panel of Fig. 4c, regions with particularly small correlation coefficients were identified in the reservoir states corresponding to certain wavenumber regions, and these correspond to the peak portions of the Raman spectra of the pMBA, as shown in the upper panel of Fig. 4c. This indicates that the diversity of reservoir states in the subject FM-RC originates from the independent nonlinear behavior of the peaks, including minor structures, as shown in Fig. 4e. The average correlation coefficient $\bar{r}_i$ (Eq. 5) for a given node $i$, sorted in decreasing order and plotted by normalized node number, is shown in Figure 4f.

$$\bar{r}_i = \frac{1}{N-1}\sum_{j \neq i}^{N} r_{ij} \quad (5)$$

The redox-IGR results are also shown as a comparison [22], with the average value $\bar{r} = \sum_{i=1}^{N}\bar{r}_i/N$ for all nodes, indicated by the dotted line. The all-node average correlation coefficients for both are almost identical, indicating that the subject FM-RC has the same high dimensionality as a redox-IGR with virtual nodes, even though it does not use virtual nodes. $\bar{r}_i$ of the FM-RC is particularly small in the region where the normalized node number is below 0.2 (that is, 20% of all reservoir states) These reservoir states correspond to the peaks of the Raman spectra. Although this study focuses on the computational power realized by only a few pMBA molecules, the coexistence of molecules that peak at different wavenumbers from pMBA, for example, has the potential to dramatically increase the overall high-dimensionality.

From these results, it can be concluded that the high computational performance of the subject FM-RC is based on good nonlinear characteristics, as shown in Figs. 4a and b, in addition to a sufficiently high dimensionality that is comparable to that of redox-IGR. Thus, it outperforms other physical reservoirs in the nonlinear waveform transformation task, where the nonlinearities outlined in Fig. 2 are particularly required for the reservoir. This is an extraordinary outcome for a result realized from only a few molecules. Furthermore, such nonlinearity and high dimensionality can be dramatically enhanced by using molecules that respond differently to ion gating in a composite manner, so the results of this study expand the possibilities of molecular computing and provide the potential for realizing new information processing systems based on the local pH response dynamics of molecules.

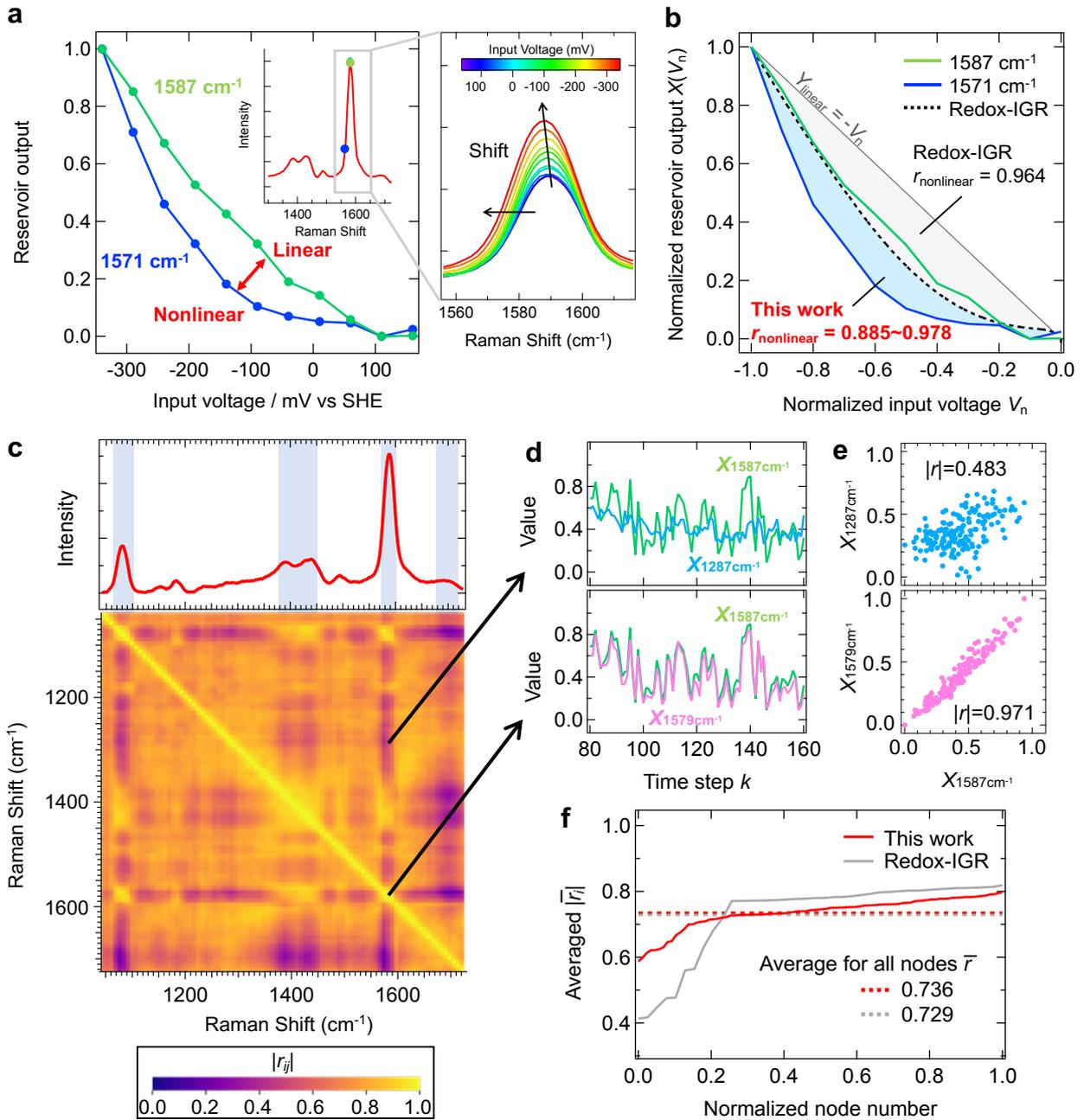

**Fig. 4 Nonlinearity and high dimensionality analysis. a** Input voltage dependence of reservoir states obtained from 1571 cm$^{-1}$ (blue line) and 1587 cm$^{-1}$ (green line). **b** Comparison of nonlinearity with Redox-IGR based on correlation coefficients. **c** Example of a Raman spectrum (upper panel) and the correlation coefficients between reservoir states for random wave inputs obtained from the Raman spectrum (bottom panel). **d** Examples of reservoir states for combinations with relatively low correlation coefficients (top) and high correlation coefficients (bottom), (**e**) and their scatter plots. **f** Comparison of averaged correlation coefficients with redox-IGR for normalized node numbers. The dashed lines show averaged values for all nodes.

## Conclusion

Few-molecule reservoir computing was firstly demonstrated on the basis of the molecular vibration dynamics of pMBA in WO$_x$@Ag-NPs systems, measured by means of SERS. The Raman signal of the few pMBA molecules, absorbed at the SERS active site of WO$_x$@Ag-NPs, were successfully detected under various local pH conditions, which were attained by applied ion-gating stimuli corresponding to input signals in RC. The reversible nonlinear response of pMBA SER spectra were utilized for performing RC on pattern recognition and prediction tasks. In nonlinear waveform transformation tasks, the subject system achieved 95.1% to 97.7% accuracy with respect to various waveforms and phase shifts, which accuracy is better than the scores reported for atomic switch nanowire network reservoirs and a solid electrolyte/diamond-based EDL-IGR [8,16,23]. In solving a second order nonlinear dynamic equation task, the system achieved 94.3% accuracy, which is comparable to other physical reservoirs reported to date [5,11,21,22]. Molecular computing has long been a goal in the study of nanoelectronics, as it is expected to enable extremely high integration due to the small volume of molecules it utilizes [44-46,59]. Since FM-RC is an effective approach to maximizing the value of the inherent properties (information) of molecules by applying the spatiotemporal dynamics of molecules directly to computations, it can be viewed as an emerging building block of the next generation of molecular computing-based AI systems, not only in conventional logic circuits for von Neumann computing, but also as reservoir computing circuits in unconventional computing [2,3]. FM-RC is particularly suited to be combined with electrochemical processes [60-64] or nanoarchitectonic materials [65-69] so as to enhance complexity as a dynamical system. Explorations are underway to identify the optimal molecules and detection systems that are most suitable for high performance molecular RC.

## Method

**Fabrication and measurement of the Few-molecule RC**

The WO$_x$ nanorods used as SERS active nanomaterials were prepared in an ultrahigh vacuum chamber at an atmosphere of $8 \times 10^{-4}$ Pa O$_2$, by sublimation of tungsten oxide from a tungsten filament. This was then heated to 1100 °C and placed 20 mm in front of a 0.2 mm diameter tungsten wire, which was heated to 700°C to 800°C [49-51]. The typical growth time was 10 hours. Numerous WO$_x$ nanorods, with diameters of 20—100 nm and lengths of 1—10 μm, were grown on the tungsten wire. One such nanorod was selected and fixed to the apex of the tungsten tip and sharpened by electrochemical etching. The WO$_x$ nanorod was fine-tuned in position using a piezo-controlled stage and fixed using epoxy resin, while being observed under an optical microscope in air.

SERS measurements were performed using a micro-Raman system (HORIBA Jobin-Yvon, HR-800). The measurements were performed in 0.15 M NaCl aqueous solution. A 488 nm wavelength laser beam was focused, with a spot diameter of 500 nm, on the WO$_x$ nanorod at the apex of a tungsten tip using a 100× water immersion objective lens. The backscattered light was collected by said objective and introduced into the spectrometer, after removing Rayleigh light with an edge filter, and was detected by a liquid nitrogen cooled CCD. The electrode potential was controlled by a potentiostat in a three-electrode configuration. A tungsten wire with a $WO_x$ nanorod attached to its apex was used as the working electrode, and platinum wires were used as the counter electrode and pseudo-reference electrode.

The subject $WO_x$ nanorod was modified with silver nanoparticles in the same configuration as the SERS measurements. 10 nM $AgNO_3$ and 6 μM poly dopamine in 0.15 M NaCl solution was used. Poly dopamine was added to confirm from the Raman spectra that the SERS active sites were formed by Ag electrodeposition. $WO_x$ nanorods were irradiated with 488 nm laser light at 2 mW intensity, while -340mV to -840mV vs SHE was applied to form silver nanoparticles, by electrodeposition of silver, at the light irradiated position. The tungsten tip, with silver-modified $WO_x$ nanorod, was removed from the aqueous solution and transferred to a 10 μM pMBA in 0.15 M NaCl aqueous solution for reservoir computing.

The silver-modified $WO_x$ nanorod was observed with TEM and EDS (JEOL, JEM-ARM200F). For TEM observation, a tungsten tip with a silver-modified $WO_x$ nanorod attached at the apex was cut into a 1 mm long piece and fixed to a TEM grid using silver epoxy.

**Nonlinear waveform transformation task**

The following explains the method employed to perform the waveform transformation task that was demonstrated to evaluate the nonlinear transformation capability of the subject molecule reservoir. In the waveform transformation task, a triangular wave was used as input, and the target waveforms $T(k)$ were sin wave (Sin), square wave (Square), $\pi/2$ phase-shifted triangular wave (Shift), and frequency-doubled triangular wave(2f). The input triangular wave was converted to a step-like voltage signal (10 s/step, 20 step/period) and input to the molecule reservoir. The applied voltage ranged from -160 mV to +340 mV vs SHE. Fourteen (14) periods of triangular waveforms were input to the molecule reservoir, and 10 periods were used for computing; excluding the first 4 periods to make the reservoir steady state. SERS measurements were made by integrating for 10 seconds corresponding to each voltage step, and the obtained Raman spectra were smoothed by moving average. From the Raman spectra in the region from 945 $cm^{-1}$ to 1728 $cm^{-1}$, the signal intensity every 3.92 $cm^{-1}$ was taken as the reservoir state. Therefore, the total reservoir size $N$ was 200. The reservoir output $y(k)$ is the linear sum of the readout weight vector $W$ and the reservoir state vector $x(k)$ at a certain time step $k$, as shown in Eq. 1. The weight matrix $W$ was learned by ridge regression, so that the reservoir output matches the target waveform $T(k)$. The cost function $J$ in the ridge regression is defined by following equation (6)

$$J(W) = \frac{1}{2}\sum_{k=1}^{L}\left(T(k) - Y(k)\right)^2 + \frac{\lambda}{2}\sum_{i=1}^{N} w_i^2 \qquad (6),$$

where $L$ (=200) and $\lambda$ (=$10^{-6}$) are the data length and regularization parameter, respectively. The trained weight vector $W$ which minimizes $J$ is given by following Eq. 7

$$\boldsymbol{W} = \boldsymbol{TX}^T(\boldsymbol{XX}^T + \lambda\boldsymbol{I})^{-1} \qquad (7),$$

where $\boldsymbol{T} = (\boldsymbol{T}(1), \boldsymbol{T}(2), ..., \boldsymbol{T}(L))$, $\boldsymbol{X} = (\boldsymbol{X}(1), \boldsymbol{X}(2), ..., \boldsymbol{x}(L))$ and $\boldsymbol{I}(\subseteq \mathbb{R}^{N \times N})$ are the target output matrix, the reservoir state matrix and the identify matrix, respectively. The computational performance of the reservoir was evaluated by the accuracy (Eq. 8) calculated from the normalized mean squared error (NMSE) between the reservoir output and the target [8,16,23].

$$\text{Accuracy} = 1 - \sqrt{\text{NMSE}} \tag{8}$$

$$\text{NMSE} = \frac{1}{L} \frac{\sum_{k=1}^{L}(T(k) - y(k))^2}{\sum_{k=1}^{L} T^2(k)} \tag{9}$$

**Solving a second-order nonlinear dynamic equation task**

In the second-order nonlinear equation task, a random input $u(k)$ was converted to a step-like voltage signal from -160 mV to +340 mV vs SHE and input to the device. As in the waveform conversion task, one step was 10 seconds, and the Raman spectra were measured by integrating for 10 seconds, corresponding to each voltage step. The input random wave was 255 steps. To avoid the influence of noise and to verify the computational capability of the molecule reservoir, the input was applied six times and the average of the Raman spectra obtained was smoothed by a moving average. From the Raman spectra in a region of from 1041 cm$^{-1}$ to 1726 cm$^{-1}$, the signal intensity every 7.87 cm$^{-1}$ was taken as the reservoir state. Therefore, the total reservoir size $N$ was 88. Reservoir states were normalized from 0 to 1, and the readout weights were trained by ridge regression as in the waveform transformation task. However, the ridge parameter λ was 1.0 and the data length $L$ was 150 for the training phase and 100 for the testing phase. Computational performance was evaluated by the accuracy shown in Eq. 8. In a previous study [5,11,21,22], to which we compared the results as shown in Fig. 3d, the computational performance was reported by NMSE as shown in Eq. 9, so this was converted to the accuracy in Eq. 8 and subsequently presented.

**Acknowledgement:**

This work was in part supported by Japan Society for the Promotion of Science (JSPS) KAKENHI Grant Number JP22H04625 (Grant-in-Aid for Scientific Research on Innovative Areas "Interface Ionics"), and JP22KJ2799 (Grant-in-Aid for JSPS Fellows). A part of this work was supported by the Iketani Science and Technology Foundation.

**Author contributions**:

D.N., Y.S., T.T., and K.T. conceived the idea for the study. D.N., Y.S. and T.T. designed the experiments. D.N., Y.S. and T.T. wrote the paper. D.N. and Y.S. carried out the experiments. Y.S. prepared the samples. D.N., Y.S. and T.T. analyzed the data. All authors discussed the results and commented on the manuscript. K.T. directed the projects.